\documentstyle[12pt]{article}

\def\thefootnote{\fnsymbol{footnote}}
\renewcommand{\thefootnote}{\alph{footnote}}

\newcommand{\rref}[1]{(\ref{#1})}
\newcommand{\beqn}{\begin{equation}}
\newcommand{\eeqn}{\end{equation}}
\newcommand{\beqarr}{\begin{eqnarray}}
\newcommand{\eeqarr}{\end{eqnarray}}

\newcommand{\matc}{\begin{array}{c}}
\newcommand{\matcc}{\begin{array}{cc}}
\newcommand{\matccc}{\begin{array}{ccc}}
\newcommand{\matcccc}{\begin{array}{cccc}}
\newcommand{\emat}{\end{array}}

\begin{document}

\begin{titlepage}

November 1998         \hfill
\begin{center}
\hfill    UCB-PTH-98/51 \\
\hfill    LBNL-42423	 \\

\vskip .15in
\renewcommand{\thefootnote}{\fnsymbol{footnote}}
{\large \bf A Note on the BPS Spectrum of the
Matrix Model}\footnote{This 
work was supported in
part by the Director, Office of Energy Research, Office of High Energy and 
Nuclear Physics, Division of High Energy Physics of the U.S. Department of 
Energy under Contract DE-AC03-76SF00098 and in part by the National Science
Foundation under grant PHY-95-14797}
\vskip .3in

Daniel Brace\footnote{email address: brace@thwk2.lbl.gov} and
Bogdan Morariu\footnote{email address: morariu@thsrv.lbl.gov}  
\vskip .3in

{\em 	Department of Physics  \\
	University of California   \\
				and	\\
	Theoretical Physics Group   \\
	Lawrence Berkeley National Laboratory  \\
	University of California   \\
	Berkeley, California 94720}
\end{center}
\vskip .3in

\begin{abstract}
We calculate, using noncommutative supersymmetric Yang-Mills gauge theory, 
the part of the spectrum of the toroidally compactified Matrix theory 
which corresponds to quantized electric fluxes on two and three tori. 
\end{abstract}
\end{titlepage}

\newpage
\renewcommand{\thepage}{\arabic{page}}

\setcounter{page}{1}
\setcounter{footnote}{0}
In this note we investigate the part of the spectrum corresponding to 
electric fluxes of
the noncommutative supersymmetric Yang-Mills (NCSYM) gauge theory~\cite{CR}
compactified on a torus. This
gives a description of the DLCQ of M-theory~\cite{LS} compactified 
on a dual torus.
Since the 
spectrum is invariant under the T-duality group $O(d,d\,|{\bf Z})$,
where $d$\, is the dimension of the compactification torus,
we can first calculate the spectrum in the simplest case which corresponds
to a NCSYM gauge theory on a
trivial bundle. Then we can use a duality transformation to rewrite 
the result in terms of the defining parameters of a dual theory
on a nontrivial bundle. 
We will also obtain this result
directly by quantizing the free system of 
collective coordinates of the twisted $U(n)$ theory. 

To obtain the spectrum we mod out gauge equivalent configurations and show
that the zero modes of the gauge field live on a torus. 
In the classical case one can find a global 
gauge transformation whose sole effect is
a shift in the zero mode of the gauge field. Then
the electric fluxes which are the 
conjugate variables are integrally quantized.
However, for a nonvanishing deformation parameter the gauge transformation
also results in a finite translation~\cite{CDS,HV,CH}. Then, just as for the
electric charge of dyons~\cite{EW}, the electric flux spectrum, for 
states carrying momentum, contains an additional
 term proportional to the deformation parameter. 
We obtain a spectrum in agreement with similar calculations in the 
literature~\cite{CDS,OPR,PMH,OP,AAS,HV,CH}.
 
Throughout this paper we will use the same notation as in~\cite{BMZ}
where it was shown explicitly that the action of 
a $U(n)$ NCSYM, with magnetic fluxes $M^{ij}$, can be written as 
the action of a $U(q)$ NCSYM on a trivial quantum bundle, 
where $q$ is the greatest common divisor of $n$ and $M^{ij}$.
Thus these two theories must have identical spectra. 
The noncommutative pure gauge theory action was written first in~\cite{CR}.
We can then introduce additional fields such that we obtain
the maximally supersymmetric $U(n)$ NCSYM gauge theory~\cite{CDS,AS2} 
\[
{\cal S}^{U(n)} = \frac{1}{g^{2}_{SYM}} \int dt
\int d^{d} \sigma~\sqrt{\det(G^{kl})}
~{\rm tr} \left(
\frac{1}{2} G_{ij} {\cal F}^{0i}  {\cal F}^{0j} -  \right.
\]
\beqn
\frac{1}{4} G_{ij} G_{kl} ({\cal F}^{ik}-{\cal F}^{ik}_{(0)})
({\cal F}^{jl}-{\cal F}^{jl}_{(0)}) +
\label{Unaction}
\eeqn
\[
\frac{1}{2} \sum_{a} \dot{X}^a  \dot{X}^a -
\frac{1}{2} \sum_{a} G_{ij} [D^i,X^a] [D^j,X^a]+
\]
\[
\left.
\frac{1}{4} \sum_{a,b} [X^a,X^b] [X^a,X^b]
+{\rm fermions}\right).
\]
As in~\cite{CDS,AS2} the action contains magnetic backgrounds
which we chose as in~\cite{BMZ}, so that the vacuum energy vanishes.

In general two NCSYM theories are dual to each other if there exists an
element $\Lambda$ of the duality group 
$SO(d,d\,|{\bf Z})$ with the block decomposition\footnote{The 
$SO(d,d\,|{\bf Z})$ subgroup of the T-duality group $O(d,d\,|{\bf Z})$ is
the subgroup that does not exchange Type IIA and IIB string theories.}
\beqn
\Lambda=
\left(
\begin{array}{cc}
{\cal A} & {\cal B} \\
{\cal C} &{\cal  D}
\end{array}
\right),
\label{lamb}
\eeqn
and a corresponding
Weyl spinor representation matrix $S$, such that their defining parameters
are related as follows
\beqarr
\bar{\Theta}
&=&
 ({\cal A}\Theta +{\cal B})({\cal C}\Theta+{\cal D})^{-1},
\label{Theta2}  
\\
\left(
\begin{array}{c}
\bar{n}\\
\bar{M}^{23}\\
\bar{M}^{31}\\
\bar{M}^{12}\\
\end{array}
\right)
&=&
S
\left(
\begin{array}{c}
n\\
M^{23}\\
M^{31}\\
M^{12}\\
\end{array}
\right),
\label{spinor}
\\
\bar{g}_{SYM}^{2}  
&=& 
\sqrt{\,|\det({\cal C}\Theta+{\cal D})|}~g_{SYM}^{2}.
\label{gsym}
\\
 \bar{G}^{ij} 
&=& 
({\cal C}\Theta +{\cal D})^{i}_{\,k} 
({\cal C}\Theta +{\cal D})^{j}_{\,\,l} \, G^{kl},
\label{dual}  
\eeqarr
Equation~\rref{spinor} was written for the three dimensional case.  
Equation~\rref{Theta2} was first written in~\cite{AS1}. A version of
equation~\rref{spinor} appeared in~\cite{AS2} were $S$ was identified as a
canonical transformation. The transformation of the coupling 
constant~\rref{gsym} and the explicit transformation of the metric~\rref{dual}
were found in~\cite{BMZ}, where also the 
transformation~\rref{spinor} was identified as a chiral spinor transformation.

All the equations in this paper where $d$ is unspecified, 
are valid for the two and three 
dimensional case, but some may have to be modified in higher dimensions.
For simplicity we consider the case when 
$n$ and $M^{ij}$ are relatively prime.
Then one can find a duality transformation $\Lambda$
such that $\bar{n}=1$ and $\bar{M}=0$ as was shown in~\cite{BMZ}.
From this point on, when we discuss the $U(n)$ theory
we will use the the $d$-dimensional block 
matrices~\rref{lamb}, with $\Lambda$ the particular
transformation that takes the $U(n)$ theory into a $U(1)$ theory. 
For example the constant background field strength can be expressed 
in terms of the block components of $\Lambda$ as
\beqn
{\cal F}_{(0)} = \frac{-1}{2\pi}({\cal C}\Theta + {\cal D})^{-1} {\cal C}.
\label{F0}
\eeqn
We write the connection as a sum of a constant curvature
$U(1)$ connection $\nabla_i$, a zero mode $A^i_{(0)}$, and a 
fluctuating part $A^i$
\[
\nabla^i-i A_{(0)}^{i}{\bf 1}-iA^{i}(Z_j)=
\partial^i+iF^{ij}\sigma_j-i A_{(0)}^{i}{\bf 1} 
-i A^{\,i}(Z_j). 
\]
Note that $A^i$\, does not contain the zero mode. 
The $Z_i$'s are 
$n$-dimensional matrices which generate 
the algebra of adjoint sections.  
For example, in the two dimensional case we have~\cite{CR,CDS,PMH,BB,BMZ}
\[
Z_1=e^{i\sigma_i Q^{i}_1/n} V^{b},~Z_2=e^{i\sigma_i Q^{i}_2/ n} U,
\]
where $U$ and $V$ are the clock and shift matrices and $Q$ is a two dimensional
matrix which reduces to the identity in the commutative case.
Substituting this in the
action we obtain 
\[
{\cal S}^{U(n)} = \frac{1}{g^{2}_{SYM}} \int dt
\int d^{d} \sigma~\sqrt{\det(G^{kl})}
~{\rm tr} \,
\frac{1}{2} G_{ij} \partial^0 A_{(0)}^{i}  \partial^0 A_{(0)}^{j}+
\ldots ,
\]
where the dots stand for terms containing only ${A}^{\,i}$. Thus classically
the zero modes decouple, and the action is just that of a free
particle
\[
{\cal S}^{U(n)}_{(0)} = \int dt\,
\frac{(2\pi)^{2}}{2} \,{\cal M}_{ij}\,
 \dot{A}_{(0)}^{i}  \dot{A}_{(0)}^{j},
\]
where the mass matrix is given by
\beqn
{\cal M}_{ij}=|n-\frac{1}{2}{\rm tr}(M\Theta)|\, 
\frac{(2\pi)^{d-2}\,\sqrt{\det(G^{kl})}}{g_{SYM}^{2}}\,G_{ij}.
\label{mass}
\eeqn
In the commutative case the first factor on the right hand side 
of~\rref{mass} reduces to $n$ and arises from taking the trace. The
origin of this factor in the noncommutative case was discussed 
in~\cite{CDS,BMZ}.
The corresponding Hamiltonian is then\footnote{This only includes the energy 
coming form the electric zero modes.}
\beqn
{\cal H}^{U(n)}_{(0)} =\frac{1}{2} {\cal M}^{ij} E^{(0)}_i E^{(0)}_j,
\label{Ham}
\eeqn
where ${\cal M}^{ij}$ is the inverse mass matrix and 
 $E^{(0)}_i$ is the momentum conjugate to $A_{(0)}^i$
\[
E^{(0)}_i=\frac{1}{2\pi i}\,\frac{\partial}{\partial A_{(0)}}.
\]
Note that $E^{(0)}_i$
correspond to zero modes of the electric field.

Before calculating the spectrum of~\rref{Ham} directly, we will use 
the duality invariance of the spectrum and obtain it by using
the simpler dual $U(1)$ theory.
We will use primes for
all the variables in the $U(1)$ theory. 
In this case the mass matrix takes the form
\beqn
{\cal M}'_{ij}= 
\frac{(2\pi)^{d-2}\,\sqrt{\det({G'}^{kl})}\,}{{g'}_{SYM}^{2}}\,G'_{ij}.
\label{mass1}
\eeqn
Just as in the commutative $U(1)$ supersymmetric 
gauge theory~\cite{W} the zero modes live on a torus. 
To see this consider the gauge transformations\footnote{We remind the reader
that we use the notation of~\cite{BMZ} were the definition of $U_i$
differs from~\cite{CDS}}
\[
U'_i=e^{i \sigma'_i}.
\]
These gauge transformations are single valued and leave the trivial
transition functions invariant.
Under these gauge transformations the connection transforms as
\[
 U'^{-1}_j ({\partial'}^i - i{A'}_{(0)}^i-i A'^i(U'_k)) U'_j 
= {\partial'}^i - i({A'}_{(0)}^i-{\delta}^i_j)-
i A'^i(e^{-2\pi i \Theta'_{jk}}U'_k).
\]

For vanishing $\Theta'$ the effect of these gauge transformations is just 
a shift of the zero mode and we have the following
gauge equivalences 
${A'}_{(0)}^i \sim {A'}_{(0)}^i +\delta^i_j$. 
Note that 
$\delta^i_{(j)}$\, for $j=1,\ldots,d$\, form a basis for a lattice $L'$
and the configuration space is ${\bf R}^{d}/{L'}$.
The conjugate momenta are then quantized
\[
E'^{(0)}_i =  n'_i,
\]
and the spectrum of zero modes is then given by
\[
{\cal E}^{U(1)} =\frac{1}{2}
{{\cal M}'}^{ij} n'_i n'_j.
\]
However in the noncommutative case 
we see that the above gauge transformations  
also produces a translation
in the $k$ direction proportional 
to $\Theta'_{jk}$. This results in a modification of the spectrum 
similar to the Witten-Olive effect~\cite{EW}. 
Let as define the total
momentum operator
operators $P'_i$ such that
\beqn
[P'^{i} , \Psi]=-i\, \partial'^{i} \Psi,
\label{Pprime}
\eeqn
where $\Psi$ is an arbitrary field of the theory.
The momentum $P'^{i}$ defined by~\rref{Pprime}
is not the standard gauge invariant 
total momentum but the difference between the two is the generator of 
a gauge 
transformation with the gauge parameter equal to the $i$-component 
of the gauge field. Thus on gauge invariant states the total momentum
defined above and the gauge invariant momentum have the same effect.

The operator generating the gauge transformation is~\cite{HV,CH}
\beqn
\exp(2\pi i (E'^{(0)}_j+\Theta'_{jk} P'^{k})).
\label{Equant}
\eeqn
Translation by an integral number of periods 
on a trivial bundle must leave the physical system invariant. The
operators generating these translations are given by
\beqn
\exp(2\pi i P'^{k}). \label{Pquant}
\eeqn
The operators~\rref{Equant} and \rref{Pquant} act as the identity 
on physical states
so  we obtain the quantization
\[
E'^{(0)}_j+\Theta'_{jk} P'^{k} = n'_j,~~ P'^{j}=m'^j,
\]
where $n_j$ and $m^j$ are integers. The spectrum of zero 
modes is then given by
\[
{\cal E}^{U(1)} =\frac{1}{2}
{{\cal M}'}^{ij} (n'_i-\Theta'_{ik} m'^k)(n'_j-\Theta'_{jl} m'^l).
\]
We can describe this result in geometric terms. In the sectors
of nonvanishing momentum the wave 
function for the zero modes is not strictly speaking 
a function but rather a section on
a twisted bundle over the torus ${\bf R}^d / L'$ with twists 
given by $\exp\,(\Theta'_{ik} m'^{k})$.

Using the duality transformations~\rref{dual} we can express 
the spectrum in terms of the $U(n)$ parameters
\beqn
{\cal E}^{U(n)} =\frac{1}{2}\,
\frac{g^{2}_{SYM}}{(2\pi)^{d-2} 
\sqrt{\det(G^{ij})}}|\det({\cal C}\Theta+{\cal D})|^{-1/2}\,
\label{result}
\eeqn
\[
\times
G^{ij} (n_i-\Theta_{ik} m^k)
( n_j- \Theta_{jk} m^k),
\]
where we aslo performed a duality transformation on the quantum 
numbers~\cite{HV,CH} 
\beqn
\left(
\begin{array}{c} n'_i \\ m'^i \end{array}
\right)
=
\left(
\begin{array}{cc}
 {\cal A} &{\cal B}\\ 
 {\cal C} &{\cal D}
\end{array}
\right)
\left(
\begin{array}{c} n_i \\ m^i \end{array}
\right).
\eeqn
Next we consider in more detail the two dimensional case. The parameters 
of the $U(1)$ and $U(n)$ NCSYM are related by the 
$SO(2,2\,|{\bf Z})$ transformation~\cite{BMZ}
\beqn
\Lambda =\left(
\matcc
a  I_2 & b \varepsilon \\
-m \varepsilon & n I_2
\emat
\right), \label{lamtwo}
\label{nmdual}
\eeqn
where $\varepsilon$ is a two dimensional matrix with the only
nonvanishing  
entries given by 
$\varepsilon_{12}= -\varepsilon_{21}=1$.
In this case
$({\cal C}\Theta +{\cal D})^{i}_{\,j} = (n + \theta m) \, \delta^i_j$
and the spectrum is
\[
{\cal E}^{U(n)} =\frac{1}{2}\,
\frac{g^{2}_{SYM}}{(2\pi)^{d-2}  \, |n + \theta m|
\sqrt{\det(G^{kl})} }\,
G^{ij} (n_i+\theta m_i)(n_j +\theta m_j),
\]
where $m^i = \varepsilon^{ij} m_j$.
This result\footnote{Expressed in terms of the string coupling 
constant of the auxiliary string theory the spectrum takes the simpler form 
$E^{U(n)}=
\frac{1}{2}{g_{s}}{| n + \theta m|^{-1}} \,G^{kl} (n_k+\theta m_k)
( n_l + \theta m_l)$.} has the expected 
factor of $|n + \theta m|$ in the denominator. 
In the DLCQ formulation of M theory this factor is proportional to the kinetic 
momentum in the compact light-like direction and is expected to appear
in the denominator of the DLCQ Hamiltonian.

Next we will obtain the spectrum directly in the $U(n)$ theory. We will do 
this in two ways. First,  consider the generators of the adjoint algebra, 
the $Z_i$'s. These generators satisfy
\beqn
Z_k(\sigma_i +2\pi\delta^j_i)=\Omega_j Z_k(\sigma_i)\Omega_j^{-1} .
\label{Z}
\eeqn
Besides having the privileged role of generators 
for the sections of the adjoint bundle,
the $Z_i$'s can also be used to perform gauge 
transformations since they are unitary. We can rewrite~\rref{Z} as
\beqn
\Omega_j=Z_k(\sigma_i +2\pi\delta^j_i)^{-1}\Omega_j Z_k(\sigma_i).
\label{OZ}
\eeqn
The right hand side of~\rref{OZ} gives the  transformation of the
transition 
functions under the $Z_i$ gauge transformation. We see that, 
just as in the $U(1)$ case where the gauge transformations $U'_i$\, left 
the transition functions trivial,
the $Z_i$'s leave the transition functions invariant.
Following the same strategy as in the $U(1)$ case, where 
we  used the $U'_i$\, to find the configuration space of the zero modes,
we can use here $Z_i$
\[
Z^{-1}_j (\nabla^i- i A^i_{(0)}{\bf 1} - i A^i(Z_k) ) Z_j =
\]
\[
\nabla^i - i (A^i_{(0)}-(({\cal C}\Theta + {\cal D})^{-1})^i_{\,j}){\bf 1}
-i  A^i(e^{-2\pi i \Theta'_{jk}}Z_k).
\]
Note that again we have separated the zero mode of the gauge connection
and we have used the identity~\cite{BMZ}  
\[
[\nabla^i, Z_j] = i(({\cal C}\Theta + {\cal D})^{-1})^i_{\,j}\, Z_j.
\]
One can express the gauge transformed connection as
\beqn
e^{-2\pi  ({\cal A}\Theta+{\cal B})_{jk} \nabla^k}
\left(
\nabla^i-i A^{i}_{(0)}{\bf 1} -i  A^i(Z_k)
\right)
e^{2\pi  ({\cal A}\Theta+{\cal B})_{jk} \nabla^k}
+i{\cal A}^{\,\,i}_{j}{\bf 1},
\label{shift}
\eeqn
where we used
\[
({\cal C}\Theta+{\cal D})^{-1}=
({\cal A}-\Theta'{\cal C})^{T}
\]
and~\rref{F0} to rewrite the extra shift in the zero mode.

Next we define the momentum operator by its action on the fields of
the theory. For example on the gauge fields $P^i$ acts as
\beqn
[P^i, A^{j}_{(0)}{\bf 1}+A^{j}(Z_k)]= -i [\nabla^i,A^{j}(Z_k)]
-i {\cal F}^{ij}_{(0)}. \label{noncovP}
\eeqn
Note that $P^i$ also acts on the zero mode $A_{(0)}^i$. This can be understood
as follows. When we define the momentum we have the choice whether to
include as part of the system the magnetic background ${\cal F}^{ij}_{(0)}$.
The standard gauge invariant momentum for which the momentum density is
${\rm tr}(F^{ij}E_j)$ can be written as the sum of two terms.
The first is just the momentum translating
the part of the system that does not include the magnetic background and
whose momentum density  is ${\rm tr}((F^{ij}-F^{ij}_{(0)})E_j)$.
The second term is an operator shifting
the zero mode of the gauge field as in~\rref{noncovP}.
Then our $P^i$ can be identified, up to the generator of a gauge 
transformation,
with the total momentum that includes the magnetic background. 
Furthermore,  we can identify,
up to the generator of a gauge 
transformation,
the first term on the right hand side of~\rref{noncovP} as the action of the
momentum operator that translates only the fluctuating part.
As we will see later it is the momentum whose density
is ${\rm tr}(F^{ij}E_j)$ that appears in the
$SO(d,d\,|{\bf Z})$ duality transformation.

A convenient way of writing the action of $P^i$ on the gauge field is
\[
[P^i, -i A^{j}_{(0)}{\bf 1}-i A^{j}(Z_k)]=
-i [\nabla^i,\nabla^j-i A^{j}_{(0)}{\bf 1}-iA^{j}(Z_k)].
\]
Then using~\rref{shift} we see that
the quantum operator which implements the gauge  transformation above 
is given by
\beqn
\exp \left(2\pi i({\cal A}_j^{\,\,k}(E^{(0)}_k+\Theta_{kl}P^l) +
   {\cal B}_{jk}P^k) \right).
\label{Zconstraint}
\eeqn
The momentum operator $P^{i}$ has integer eigenvalues since the space is a 
torus with lengths $2\pi$. 
One can also see this by considering the operator
\beqn
\exp \left(2\pi i({\cal C}^{ji}E^{(0)}_i +
   ({\cal C}^{jk}\Theta_{ki} +  {\cal D}^j_{\, \, i})P^i) \right).
\label{constraint}
\eeqn
This acts trivially on every operator in the $U(n)$ theory. 
In particular the
combination of operators in the exponent has no effect on the zero mode.
The condition that~\rref{Zconstraint} and~\rref{constraint}
should act as the identity on the physical Hilbert space is equivalent to
the quantization
\beqarr
{\cal A}_j^{\,\,k}(E^{(0)}_k+\Theta_{kl}P^l) +
   {\cal B}_{jk}P^k &=& n'_j,  \nonumber \\
{\cal C}^{jk}(E^{(0)}_k +
  \Theta_{kl} P^{l}) +  {\cal D}^j_{\, \, k}P^k &=&m'^j.\nonumber
\eeqarr
Since the matrices ${\cal A, B, C, {\rm and}~ D}$ are the block components
of an element of 
$SO(d,d|{\bf Z})$ this is equivalent to
\[
E^{(0)}_j+\Theta_{jk} P^{k} = n_j,~~ P^{j}=m^j,
\]
where $ n_j$ and $m^j$ are integers.
Using the Hamiltonian~\rref{Ham} and the above quantization the electric
flux spectrum is
\[
{\cal E}^{U(n)} =\frac{1}{2}\,
\frac{g^{2}_{SYM}}{(2\pi)^{d-2} 
\sqrt{\det(G^{ij})}}|n - \frac{1}{2} {\rm tr}(M \Theta)|^{-1}\,
\]
\beqn
\times
G^{ij} (n_i-\Theta_{ik} m^k)
( n_j- \Theta_{jk} m^k),
\label{Unspec}
\eeqn
which is identical to the result~\rref{result} obtained by duality.

Finally we present an alternative derivation of the spectrum using
the gauge transformations $\exp(i\bar{\sigma}_i)$ 
where $\bar{\sigma}_i= \sigma_j Q^j_{\,\,i}$ and $Q^j_{\,\,i}$ is a matrix
defined in~\cite{BMZ} and equals the identity for vanishing
deformation parameter or magnetic background. 
This derivation is closely related
to the derivation of the spectrum in~\cite{HV,CH}.  As discussed 
in~\cite{BMZ} gauge invariant quantities such as the Lagrangian density
have periodicity $2\pi$ in the $\bar{\sigma}_i$ variables. Then we can use
$\bar{U}_i=\exp(i\bar{\sigma}_i)$ as a gauge transformation just as we
used $U'_i$ in the $U(1)$ theory. Note first that $\bar{U}_i$ is a globally
defined gauge transformation. It is convenient to write it as
$\bar{U}_i=
{\cal U}_i \,e^{2\pi \Theta_{ij}\nabla^j}$.
Here ${\cal U}_i= e^{i\sigma_i -2\pi \Theta_{ij}\partial^j}$ and is 
the variable implementing the 
quotient condition~\cite{CDS}.
The effect of this gauge transformation is
\[
\bar{U}_j^{-1} (\nabla^i- i A^i_{(0)}{\bf 1} - i A^i(Z_k) ) \bar{U}_j =
\]
\[
e^{-2\pi \Theta_{ij}\nabla^j} 
(\nabla^i- i A^i_{(0)}{\bf 1} - i A^i(Z_k) )
e^{2\pi\Theta_{ij}\nabla^j} +i\, \delta^i_{\,j}.
\]
The operator implementing this gauge transformation in the Hilbert 
space is
\[   
\exp \left(-2\pi i(E^{(0)}_k+\Theta_{kl}P^l) \right).
\]
Again, on gauge invariant states this operator acts trivially and 
together with the quantization of the momentum results in the same 
spectrum~\rref{Unspec} as using $Z_i$.
Note that the second method 
of deriving  the $U(n)$ spectrum is similar in spirit 
to the derivation of the $U(1)$ spectrum. For example the gauge 
transformation is an element of the $U(1)$ subgroup.
However, the first derivation is instructive since it exhibits
inside the $U(n)$ theory the dual $U(1)$ theory variables
such as $P'^i$ and $E'_i$. 

\section*{Acknowledgments}
We would like to thank
Bruno Zumino for discussions and valuable comments. This paper also benefited 
from  communications with Christiaan Hofman and Albert Schwarz.
This work was supported in part by 
the Director, Office of Energy Research, Office of High Energy and Nuclear
Physics, Division of High Energy Physics of the U.S. Department of Energy
under Contract DE-AC03-76SF00098 and in part by the National Science 
Foundation under grant PHY-95-14797.

\end{document}